\documentclass[twocolumn,showpacs,prl,amsmath,amssymb,superscriptaddress]{revtex4}
\usepackage{graphicx}
\usepackage{dcolumn}
\usepackage{bm}
\usepackage[latin1]{inputenc}
\usepackage[mathscr]{eucal}
\usepackage{epsfig}
\usepackage{rotating}
\usepackage{multirow}

\begin{document}

\title{Primary tunnel junction thermometry}

\author{Jukka P. Pekola}
\author{Tommy Holmqvist}
\author{Matthias Meschke}
\affiliation{Low Temperature Laboratory, Helsinki University of
Technology, P.O. Box 3500, 02015 TKK, Finland}



\begin{abstract}
We describe the concept and experimental demonstration of primary
thermometry based on a four probe measurement of a single tunnel
junction embedded within four arrays of junctions. We show that in
this configuration random sample
specific and environment-related errors can be avoided. This method relates temperature
directly to Boltzmann constant, which will form the basis of the
definition of temperature and realization of official
temperature scales in the future.
\end{abstract}


\maketitle Temperature is a relatively poorly known quantity in
modern metrology. It is well recognized that the way the
international temperature scale is currently realized, in particular
towards low temperatures, needs to be seriously reconsidered. It is
currently based largely on artefacts which should be replaced by
methods relating to thermodynamic temperature via Boltzmann constant $k_B$
\cite{fellmuth,casa}. Methods based on solid state tunnel junctions,
Coulomb blockade thermometry (CBT) \cite{cbt94,bergsten99} and shot
noise thermometry (SNT) \cite{spietz03,spietz06}, have both shown
great promise as $k_B$-based thermometers for metrology. However, both
of them fall short up to now, when it comes to sufficient absolute
accuracy. In case of SNT, the limitations are mainly of practical
nature, and can possibly be overcome by a careful design of the
sensor and the measurement set-up. For CBT, an uncontrolled error
source is of more fundamental concern: CBT involves a measurement of
a series connection of nominally identical junctions. The inevitable
spread in junction parameters leads, however, to an error, which can
usually be made small, but which limits the accuracy in particular
when the average junction size is small \cite{hirvi95,farhangfar97}.
In this letter we introduce and demonstrate a method,
single-junction thermometry, SJT, which combines the advantages of
basic CBT thermometry, but which avoids the parameter dispersion
induced errors altogether. We show theoretically that the errors can
then be efficiently suppressed, and demonstrate the operation in
experiment.

\begin{figure}[t]
\includegraphics[width=1.0\linewidth]{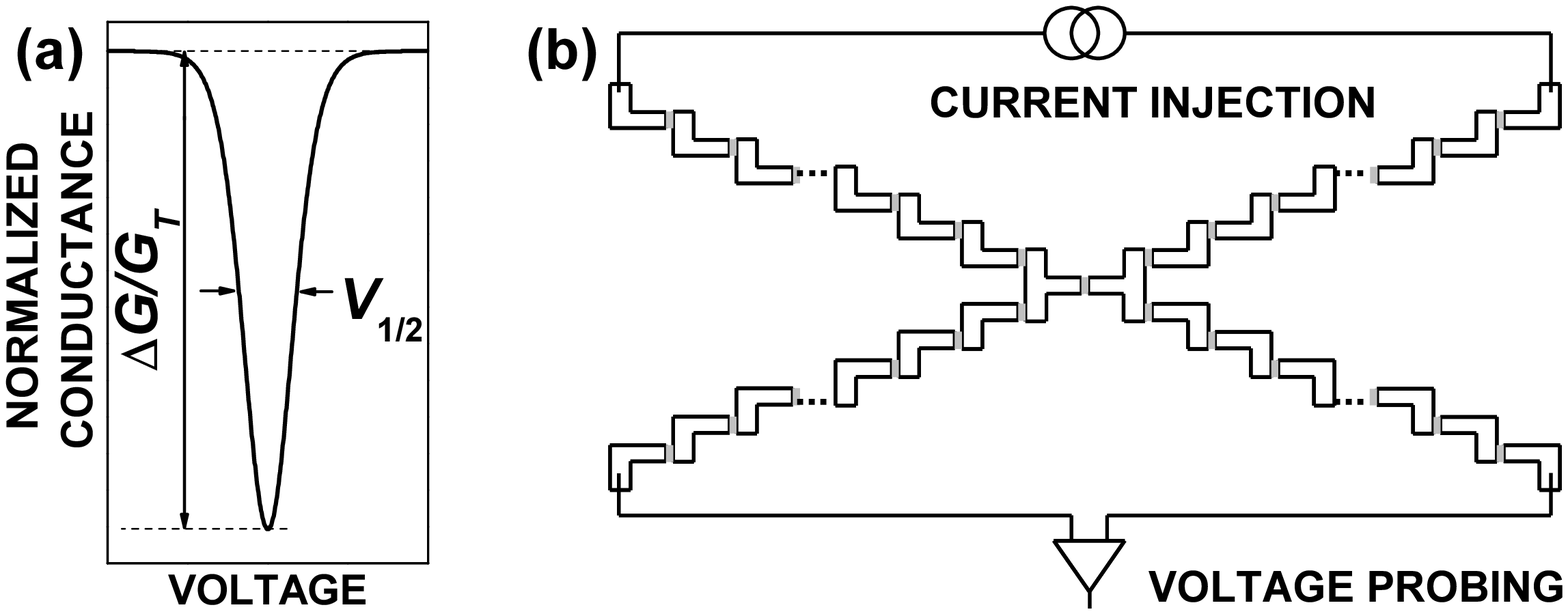}
\caption{The single junction thermometer (SJT). (a) A typical conductance curve of a thermometer. (b) Schematics of the SJT. The white bounded areas are conductors, and the grey interconnects are tunnel junctions.} \label{system}
\end{figure}

In Coulomb blockade thermometry an array of tunnel junctions shows a
drop in its differential conductance around zero bias voltage,
because of the influence of single-electron charging effects. The
ideal operation regime of a CB thermometer is determined by the
ratio of the single-electron charging energy, $E_C=e^2/2C$, where
$C$ is the (average) junction capacitance, and the thermal energy
$k_BT$ at temperature $T$ such that $E_C \ll k_BT$. The measured
conductance peak [see Fig. \ref{system}(a)] has two important characteristics, its full voltage
width at half minimum, $V_{1/2}$, and its normalized (by asymptotic
conductance at large voltages, $G_T$) depth $\Delta G/G_T$. Here the
first one is given by $5.44k_BT/e$ per junction, and serves as the primary
thermometer, provided the junctions in the sensor are mutually
identical. The latter one is inversely proportional to $T$. SJT
thermometry is based on the same principle as CBT but there the
objective is to measure the conductance of a single tunnel junction,
embedded in a four probe configuration through lines consisting of
arrays of tunnel junctions, see Fig. \ref{system}(b). In this
topology, the advantageous protection from the influence of
electromagnetic environment is achieved. At the same time, this
configuration abolishes any requirement of a uniform structure,
because only one junction is probed, and the rest of the junctions,
indeed not necessarily identical, act as an environment for this
one.

We separate the theoretical analysis into two parts. First we
consider the case where the influence of the environment beyond the
junction array can be neglected. We show that the measurement is
perfect in this case even for a non-uniform structure. Then we use
this result as the seed for analyzing the influence of the
dissipative environment on the performance of the thermometer, and
show that with sufficiently long junction arrays the accuracy can be
maintained at the desired level.

The tunnelling rate through a junction $i$ in forward ($+$) or
backward ($-$) direction with normal conductors in thermal
equilibrium is given by \cite{averin}
\begin{equation} \label{deltagamma}
\Gamma_{0,i} (\delta F_i^\pm) =\frac{1}{e^2R_{T,i}}\gamma(\delta
F_i^\pm),
\end{equation}
where $R_{T,i}$ is the junction resistance,
$\gamma(x)=x/(1-e^{-x/k_BT})$, and $\delta
F_i^\pm$ is the change in electrostatic energy in tunnelling. We may
separate this energy change as $\delta F_i^\pm = \pm eV_i+\delta
E_{{\rm ch},i}^\pm$, where $V_i$ is the (mean) voltage drop across
junction $i$ and $\delta E_{{\rm ch},i}^\pm$ is the internal energy
change associated with charging the capacitors of the array. In the
present analysis we limit ourselves to the lowest order result in
$E_C/k_BT$, which is what yields the basic results in thermometry.
In this spirit, we then expand $\Gamma_{0,i} (\delta F_i^\pm)\simeq
\Gamma_{0,i} (\pm eV_i)+\Gamma_{0,i}' (\pm eV_i)\delta E_{{\rm
ch},i}^\pm$. Analytic corrections for lower temperatures can be
obtained readily by expanding up to higher orders, but they will not
be considered here. The current $I_i$ through junction $i$ can be
obtained as $I_i =e\sum_{\{n\}} \sigma({\{n\}})[\Gamma_{0,i} (\delta
F_i^+)-\Gamma_{0,i} (\delta F_i^-)]$. Here, $\sigma({\{n\}})$ is the
occupation probability of the charge configuration ${\{n\}}$ on the
islands within the array. With these premises, and by using
identities $\gamma(x)-\gamma(-x)=x$ and $\gamma'(x)+\gamma'(-x)=1$,
we obtain
\begin{eqnarray}\label{iv1}
I_i=&&\frac{1}{eR_{T,i}}\sum_{\{n\}}
\sigma({\{n\}})[eV_i+(\delta
E_{{\rm ch},i}^+ + \delta E_{{\rm ch},i}^-)\gamma'(eV_i)\nonumber \\ &&-\delta E_{{\rm ch},i}^-].
\end{eqnarray}
The internal charging energy for each charge configuration ${\{n\}}$
is given by $E_{\rm ch}=\frac{e^2}{2}\sum_{\{n\}} ({\bold
C}^{-1})_{i,j}n_i n_j$, where ${\bold C}^{-1}$ is the inverse
capacitance matrix of the junction array. We have neglected
the offset charges on the islands since in the high temperature regime the charge distribution is quasi-continuous, $ \langle \delta n_k^2 \rangle \gg
1$ \cite{hirvi95}. Let us denote the islands surrounding the junction $i$ by L and
R. Then for the relevant processes only $n_L$ changes into $n_L\mp
1$ and $n_R$ to $n_R\pm 1$, whereas all the other charge numbers
remain constant. There are two junction
connections to islands L and R in SJT. Evaluating $\delta E_{{\rm
ch},i}^\pm$ as the difference of $E_{\rm ch}$ for the charge
configurations before and after the tunnelling event, and making use
of the properties $\sum_{\{n\}}\sigma(\{n\})=1$ and
$\sum_{\{n\}}n_k\sigma(\{n\})=0$ for all $k$ because of the symmetry
of $E_{\rm ch}$, we obtain the normalized conductance of
junction $i$, $(G/G_T)_i \equiv R_{T,i}\frac{dI_i}{dV_i}$ as
\begin{equation} \label{cond}
(\frac{G}{G_T})_i = 1-\frac{\delta_i}{k_BT}g(v_i),
\end{equation}
where $\delta_i=e^2[({\bold C}^{-1})_{LL}+({\bold
C}^{-1})_{RR}-2({\bold C}^{-1})_{LR}]$, $g(x)=e^x[e^x(x -2)+ x
+2]/(e^x-1)^3$ and $v_i=eV_i/k_BT$. The result of Eq. \eqref{cond}
is in fact the basis of the standard CBT formula in linear arrays of
junctions. Equation \eqref{cond} tells that conductance of a single
junction in SJT is accurate as a thermometer in any array of
junctions: the magnitude of conductance suppression depends on the
distribution of junction sizes
via the capacitance matrix ($\delta_i$), but the temperature can be
determined unambiguously from, e.g., the half width of $g(v_i)$, if
$V_i$ can be measured. Error-free measurement of $V_i$ is indeed
possible in the configuration of Fig. \ref{system}(b). This happens
since the voltage measurement via two arrays is typically performed using an
amplifier with very large input impedance (in any case
much larger than the resistance of the junction arrays). Then,
essentially no current flows through these two arrays, and there is
no voltage drop across them. Thus $V_i$ is indeed the voltage seen
by the amplifier.

The analysis above applies to the case where the connection to the bias sources and signal amplifiers has zero impedance. In practice this is not the case, and the
environment impedance introduces errors to Coulomb blockade
thermometry which can be suppressed by using long arrays of
junctions \cite{farhangfar97}. Similarly, one can realize the single
junction measurement which avoids the errors by the
embedding arrays. To analyze the remaining errors in this case
quantitatively, we may write the tunnelling rates $\Gamma_i^\pm$
instead of $\Gamma_{0,i}^\pm$ of Eq. \eqref{deltagamma} as
\begin{equation} \label{eq1}
\Gamma_i^\pm =\frac{1}{e^2R_{T,i}}\int \gamma(E')P(\delta F_i^\pm
-E') dE'.
\end{equation}
Here, $P(E)$ originates from the environment theory of
single-electron tunnelling \cite{pe}, and it yields the probability (density)
of electron to exchange energy $E$ when it tunnels: positive
(negative) $E$ refers to energy emission (absorption) by electron.
In the first analysis above we thus assumed $P(E)=\delta(E)$, i.e.,
the Dirac delta function. The same steps as in the ideal
dissipationless environment above lead now to
\begin{eqnarray} \label{gamma}
&&(\frac{G}{G_T})_i =1-\int [h(v_i+\frac{E}{k_BT})-h(v_i
-\frac{E}{k_BT})]P(E) dE\nonumber
\\ &&-\frac{\delta_i}{2k_BT} \int [g(v_i +\frac{E}{k_BT})+g(v_i -\frac{E}{k_BT})]P(E)
dE.
\end{eqnarray}
Here, $h(x)=e^{x}(e^x-x -1)/(e^x-1)^2$.  For $P(E)=\delta(E)$, Eq.
\eqref{gamma} reduces naturally to \eqref{cond}. Equation
\eqref{gamma} yields an easy way to evaluate the influence of
environment even in complex circuits.

In general $P(E)$ is obtained from the phase-phase correlation
function $J(t)$ with $P(E)=\frac{1}{2\pi\hbar}\int_{-\infty}^\infty
dt \exp[J(t)+\frac{i}{\hbar}Et]$ \cite{pe}. Here we assume that the array is uniform and embedded in a resistive environment with resistance
$R$. We
are interested in conductance of one junction within an array. $J(t)$ can
be written as
\begin{eqnarray} \label{jt}
&&J(t) =\pi \frac{R_{\rm eq}}{R_{\rm K}}\{[\cot (B)
(1-e^{-|\tau|})-\frac{|\tau|}{B}
\\ &&-2\sum_{k=1}^{\infty}\frac{1-e^{-k\pi|\tau|/B}}{k\pi(1-(k\pi/B)^2)}]-i
[sign (\tau)(1-e^{-|\tau|})]\}.\nonumber
\end{eqnarray}
Here $\tau = t/(R_{\rm eq}C_{\rm eq})$, $B=\hbar/(2k_{\rm B}TR_{\rm
eq}C_{\rm eq})$, $R_{\rm eq}=R/N^2$ and $C_{\rm eq}=NC$. For a linear array of $N$ junctions we have $\delta_i=\frac{N-1}{N}\frac{e^2}{C}$. The results for the corresponding SJT are identical to these upon replacing $N$ by $N'+1$ in the expressions above. Here $N'$ is the number of junctions in each of the four surrounding arrays. Above we have assumed that the stray capacitances are small.
\begin{figure}[t]
\includegraphics[width=1.0\linewidth]{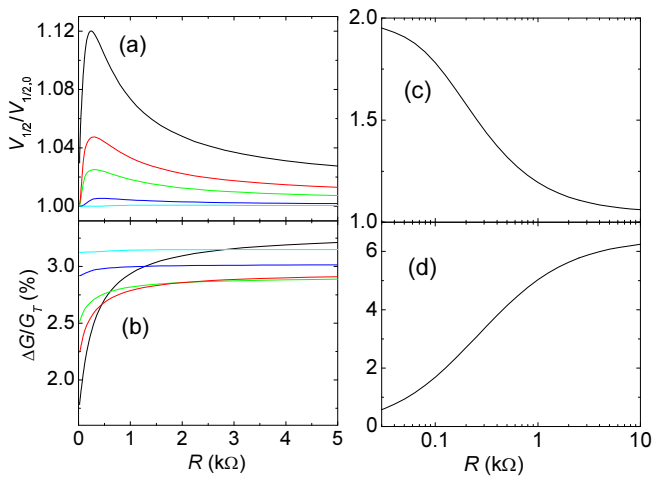}
\caption{The influence of environment on SJT thermometry. (a) $R$ dependence of the half
width of the conductance dips for $N'=1,2,3,7,15$ ($N=2,3,4,8,16$ in linear arrays)
from top to bottom.
(b) The normalized depth of the conductance drop, with the same
parameters as in (a). At low values of $R$, $N$ grows from bottom to
up. (c), (d) The corresponding quantities on the logarithmic resistance scale for a single bare junction ($N'=0$, $N=1$). In these plots $E_C/k_BT =0.1$.} \label{fig2}
\end{figure}

Figure \ref{fig2} shows numerical
results based on the analysis above. Figure \ref{fig2}(a) shows
the width as normalized to the ideal width in delta-function
environment for a junction in arrays with varying length and as a
function of $R$. It is clear that one needs to protect the junction
by a long array, if accurate measurement of temperature using $V_{1/2}$ is to be
obtained. For $N=2$, an error of about 10\% in the range $100$
$\Omega$ $\lesssim R \lesssim 500$ $\Omega$ is expected. The impedance of the environment at high frequencies is approximately $Z_{\rm env}=\sqrt{\epsilon/\mu}$,
determined by the permittivity $\epsilon$ and permeability $\mu$ of
the medium. For vacuum its value is $\simeq 377 $ $\Omega$, and for a circuit on silicon it is a few times smaller. Such increase of $V_{1/2}$ due to environment in short arrays is supported quantitatively by experiments, see, e.g., Ref. \cite{hirvi95}. The depth of the conductance
dip is shown in Fig. \ref{fig2}(b): it
depends strongly on $R$ only in short arrays. We
note that: (i) Since the environment is never known precisely in the
experiment, there is almost no way to correct theoretically for such
errors: the only working strategy then is to suppress these errors
precisely by embedding the measured junction in a long array. (ii)
The effect of error suppression is essentially proportional to
$N^{-2}$ (if $k_BTRC/\hbar < N$). Therefore, an array with $N\sim 50$ is in principle
sufficient for measurements with $10^{-4}$ absolute accuracy. (iii)
Embedding a junction in a very resistive environment \cite{joyez97}
instead of a junction array is not the best strategy in thermometry either,
which is indicated by the very slowly decaying tails of the error at
large values of $R$. To fully appreciate this point, we show in Fig. \ref{fig2}(c) and (d) the width and depth, respectively, of a single junction peak in a purely
resistive environment. Although the width at large values of $R$ slowly approaches
unity, experimentally it is hard to fabricate resistive
environments with $R \gg 10$ k$\Omega$.

\begin{figure}[t]
\includegraphics[width=1.0\linewidth]{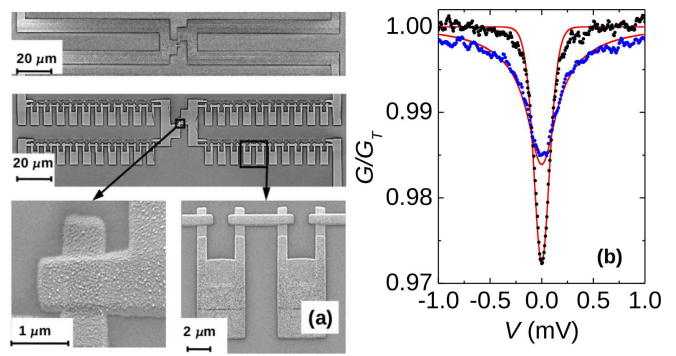}
\caption{Samples and comparison of the data on SJT and a bare single junction. (a) Electron micrographs of the reference structure with one junction connected to four leads (top), and the SJT structure with $N'=20$ junctions in the leads (centre). Zoom of the central junction and of a section of a junction array are shown at the bottom. (b) Measurement of the conductance of the bare single junction and the SJT in Sample B at $T\simeq 0.3 $ K. The deeper and narrower drop in conductance corresponds to the SJT. The calculated conductance curves for the two samples based on the model described are shown by the solid lines assuming $R=80$ $\Omega$.} \label{fig3}
\end{figure}

This concludes our proof that, theoretically, the influence of
uncontrolled error sources, the inhomogeneity of the junction array
and the noise of the environment, can be efficiently suppressed in a
SJT. We discuss next the proof-of-the-concept
experiments. Samples [see Fig. \ref{fig3}(a)] were fabricated by
electron beam patterning and shadow angle evaporation with an
oxidation step between the two electrode layers. Both the bottom and
the top electrodes are of aluminium, they are 40 nm and 45 nm thick,
respectively. The bottom electrode was thermally oxidized at 100
mbar for 10 min before deposition of the top electrode at an oblique
angle. Two samples (A and B) with two
types of structures have been measured in this work. The single tunnel junction was connected either directly to the external leads or it was
embedded within four arrays of $N'=20$ junctions, respectively. The two types of structures were fabricated on the same
chip in the same vacuum cycle. Nominally, the central junctions are identical in the two cases, and all the junctions are 0.6 $\mu$m$^{2}$ of area,
yielding a junction resistance of $\simeq 6$ k$\Omega$ (Sample A) and $\simeq 4$ k$\Omega$ (Sample B).

The samples were measured in a dilution refrigerator with a 40 mK
base temperature. However, these structures were not suitable for
very low temperature measurements: we observe strong self-heating due
to weak electron-phonon coupling in the present geometry near base temperature
\cite{meschke04}. Therefore we present here data at temperatures
at and above 150 mK. Conductance measurements in the SJT
configuration yield a deeper and narrower peak than for the unprotected single junction structure in
agreement with the calculated results of Fig. \ref{fig2}. This is
shown in Fig. \ref{fig3}(b) for Sample B. Both the SJT ($N'=20$) and the bare junction ($N'=0$) data follow the environment calculation assuming $R=80$ $\Omega$, a value which is consistent with the discussion above. The
arrays can thus indeed be employed to efficiently protect the junctions
against environment fluctuations.

\begin{figure}[t]
\includegraphics[width=1.0\linewidth]{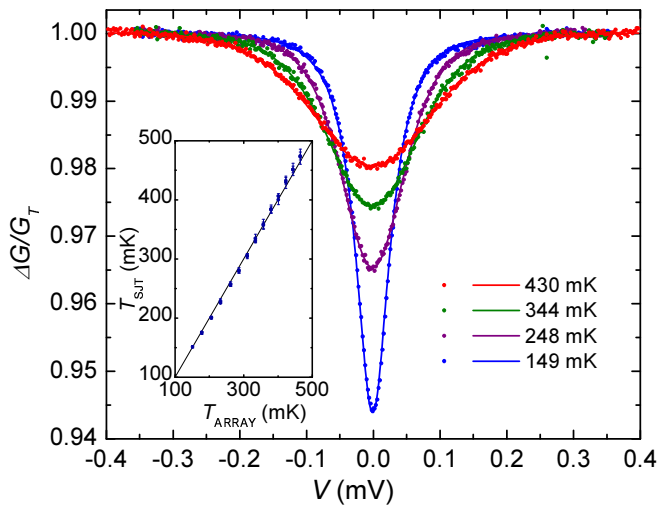}
\caption{Measurements on a SJT (Sample A). Conductance dips at four temperatures are shown together with fits according to the model presented, including the self-heating correction \cite{meschke04}. The inset shows a comparison of the temperature deduced from SJT (vertical axis) against that obtained by an ordinary CBT measurement across one of the embedding arrays (horizontal axis). The solid line has unit slope, and the error bars indicate the confidence interval of the fits.} \label{fig4}
\end{figure}

Figure \ref{fig4} shows four-probe conductance measurements of the SJT sample at a few temperatures together with fits according to the model presented. We included in the fits the influence of self-heating in the manner presented in Ref. \cite{meschke04}, and this yielded perfect match to the peaks, see the lines on top of the data. The self-heating has, however, only a small influence on the curves at the temperatures shown. The extracted temperatures from such fits are shown in the inset of Fig. \ref{fig4} against the reading of the CBT "reference thermometer", which was one of the $N'=20$ junction arrays in the same sample. The dominating discrepancies between the two thermometers are due to finite errors in the fitting procedure in each case. The agreement between the two is good over the whole temperature range in Fig. \ref{fig4}.

In the theoretical analysis we focused on the high temperature and high junction
resistance limit, and did not discuss errors due to, e.g., enhanced
Coulomb effects at lower temperatures \cite{farhangfar97} and strong
tunnelling in low resistance junctions \cite{averin92,farhangfar01}.
Yet these errors can be treated similarly to what has been done in
standard Coulomb blockade thermometry, and their influence can be
estimated and kept at a tolerable level by proper choice of junction sizes for each
temperature and with suitable tunnel barrier parameters.
One more (controllable) error to judge is the influence of the size
of the islands between the junctions, $\ell$:
as long as it is smaller than the distance to "horizon", $\ell \ll
\hbar c/k_BT$, the above lump element analysis is valid
\cite{kauppinen98,wahlgren98}. Here, $c=(\mu \epsilon)^{-1}$ is the
signal propagation speed. The condition gets critical at
particularly high temperatures, and extra care to place the array
close to the junction has to be taken then. We want to add that the
presented thermometry is not necessarily limited to the standard
planar tunnel junction design, but may be applicable, e.g., in
scanning probe or break junction geometries, since the junction parameters of the
surrounding arrays need not be the same as those of the central
junction. This might lead to the possibility of using tunable tunnel
junctions in thermometry.

Summarizing, we propose an absolute single tunnel junction
thermometer. We have demonstrated the concept in preliminary
experiments. Our method may turn out to be valuable in future
realization of the international temperature scale based on
Boltzmann constant.

We thank Mikes (Centre for Metrology
and Accreditation) and NanoSciERA project NanoFridge for financial support, Dmitri Averin for a useful discussion, and Mikko Möttönen for help in the analysis.

\end{document}